\documentstyle[12pt]{article}
 
\input epsf 

\begin{document}
 
\begin{titlepage}
 
\hspace*{\fill}\parbox[t]{2.8cm}{DESY 94-114 \\ SCIPP 94/17 \\ July 1994}
 
\vspace*{1cm}
 
\begin{center}
\large\bf
BFKL versus
${\cal O}(\alpha_s^3)$ Corrections to Large-rapidity Dijet Production
\end{center}
 
\vspace*{0.5cm}
 
\begin{center}
Vittorio Del Duca \\
Deutsches Elektronen-Synchrotron \\
DESY, D-22607 Hamburg , GERMANY\\
\vspace*{0.5cm}
and\\
\vspace*{0.5cm}
Carl R. Schmidt \footnote{Supported in part by the U.S.
Department of Energy.} \\
Santa Cruz Institute for Particle Physics\\
University of California, Santa Cruz, CA 95064, USA
\end{center}
 
\vspace*{0.5cm}
 
\begin{center}
\bf Abstract
\end{center}
 
\noindent
We examine dijet production at large rapidity intervals at Tevatron
energies by comparing an exact ${\cal O}(\alpha_s^3)$ calculation 
with the BFKL approximation, which resums the leading powers 
of the rapidity interval $y$ to all orders in $\alpha_s$. 
We analyze the dependence of the exact ${\cal O}(\alpha_s^3)$ calculation 
on the jet cone-size as a function of $y$, and use this cross section 
to define an ``effective rapidity'' $\hat y$
which reduces the error that the large-$y$ approximation induces on the
kinematics. Using $\hat y$ in the BFKL resummation, we reexamine jet 
production at large
transverse momenta and the transverse momentum decorrelation 
of the tagging jets. We find less dramatic, but still
significant, effects than found previously using the large-$y$ approximation.
 
\end{titlepage}
 
\baselineskip=0.8cm
 
\section{Introduction}
 
The state-of-the-art in jet physics at hadron colliders is described by
next-to-leading-order QCD parton-level calculations.  These consist of
the ${\cal O}(\alpha_s^2)+{\cal O}(\alpha_s^3)$ one-loop
$2\rightarrow2$ parton scattering, combined together with the
${\cal O}(\alpha_s^3)$ tree-level $2\rightarrow3$ parton
scattering\cite{ES}.
These perturbative calculations describe the hard part of the
scattering, while the nonperturbative effects are factorized
into the parton structure functions.  An advantage of going
to next-to-leading order is that it reduces the dependence on the
arbitrary scale associated with this factorization.  In addition,
the inclusion of a third final-state parton allows a more detailed
description of the jet structure.  These next-to-leading order
calculations\cite{EKS}, \cite{EKS2}, \cite{GGK} appear to be in very good
agreement with the one- and two-jet inclusive distributions obtained
from the data of the CDF experiment at the Fermilab Tevatron
Collider\cite{CDF}, \cite{Abe}.
 
Despite these successes, it is possible to imagine kinematic
configurations where this fixed-order analysis is inadequate, even
though the underlying process is still perturbative in the usual
sense.  This could occur when the cross section contains logarithms
of large ratios of kinematic invariants.  Typical invariants are
the hadron-hadron center-of-mass energy $\sqrt{s}$, the parton-parton
center-of-mass energy $\sqrt{\hat s}=\sqrt{x_Ax_Bs}$, where $x_A$ and $x_B$
are the momentum fractions of the partons originating the hard scattering,
and the momentum transfer $Q$,
which is of the order of the transverse momentum of the jets produced
in the hard scattering.
Large logarithms will appear when $\sqrt{s} \gg Q$, in the
{\it semihard region} of kinematic phase space.
We can then avoid complications associated with the
small-$x$ behavior of the structure functions by requiring
that the parton momentum fractions, $x_A$ and $x_B$, are
sufficiently large, as originally suggested by Mueller and
Navelet\cite{MN}.
In this case the large logarithms, $\ln(\hat s/Q^2)$, factorize
entirely into the partonic subprocess cross section.  These
logarithms, which are of the size of
the rapidity interval in the scattering process, can be resummed by
using the techniques of Balitsky,
Fadin, Kuraev, and Lipatov (BFKL)\cite{BFKL}.
For jet events with a large rapidity interval, the
amplitudes are dominated by contributions from multiple gluons that
uniformly fill the interval between the two extreme jets.
The BFKL theory systematically resums these leading powers in the
rapidity interval, including both real and virtual gluon corrections.
 
In a previous paper\cite{DDS} we showed how to analyze dijet production
experimentally so that it most closely resembles the configuration
assumed in the BFKL theory.  The main difference from the standard
hadronic jet analysis is that the jets are ordered first by their
rapidity rather than by their energy.  Thus, we look at all the jets
in the event
that are above a transverse momentum cutoff $p_{\perp min}$,
using some jet-definition algorithm, and rank them by their rapidity.
We then tag the two jets with the largest and smallest rapidity and
observe the distributions as a function of these two {\it tagging jets}.
The cross section is inclusive so that the distributions are affected by
the hadronic activity in the rapidity interval $y$ between the tagging
jets, whether or not these hadrons pass the jet-selection criteria.
We will refer to these hadrons in the rapidity interval as
{\it minijets}.
 
In ref.~\cite{DDS} we showed that the exponential enhancement with the
rapidity interval $y$ in dijet production at fixed $x_A$ and $x_B$ , which
was originally suggested as a signature of the BFKL minijets by
Mueller and Navelet, is highly suppressed by the parton distribution
functions at Tevatron energies.
However, other observables such as the jet transverse
momentum distribution, and the jet-jet correlations in $p_\perp$ and
azimuthal angle are significantly affected by the minijets.  For
instance, the
transverse momentum distribution was considerably enhanced at large
$p_\perp$ and large y.  This enhancement has been seen in the CDF
data\cite{Abe}, although the data analysis differs somewhat from that
needed to
compare directly with the BFKL resummation.  In addition, there should
be some dependence on the cutoff $p_{\perp min}$, and we shall see
here that the $p_\perp$
distributions should differ for the two tagging jets if they are not
symmetrically placed around
zero rapidity.  Finally, we saw that the correlation in
transverse momentum and azimuthal angle of the tagging jets is not a leading
feature of the expansion in the rapidity interval. Accordingly, it fades away
as the rapidity interval increases.  The decorrelation in
azimuthal angle has also been noted by Stirling\cite{stir}.
 
For the most part these effects at large $y$ can be easily understood
in terms of the sharing of the total $\vec{p}_\perp$ by the
additional minijets, which the BFKL resummation automatically
includes.  In a fixed-order calculation the effects arise first
at ${\cal O}(\alpha_s^3)$, where the introduction of a third
final-state parton removes some of the correlations inherent at lowest
order.  Thus, it is interesting to compare the BFKL resummation
directly with the ${\cal O}(\alpha_s^3)$ result.  This was
done for the case of the phi distribution in ref.~\cite{stir}.
Here we shall extend this comparison, which can be approached in two
different ways.  First, one can truncate the BFKL solution at
${\cal O}(\alpha_s^3)$ and compare this with the
exact ${\cal O}(\alpha_s^3)$ result.  The truncated
${\cal O}(\alpha_s^3)$ BFKL cross section is just the large $y$ limit
of the complete ${\cal O}(\alpha_s^3)$ cross section, and so this
comparison will indicate how good the leading log approximation is at
the different experiments.
Second, one can compare the full BFKL solution with its truncation to
${\cal O}(\alpha_s^3)$.  This should give an indication of the size of
the contributions from higher orders in $\alpha_s$.
In addition it will help in isolating the distributions which are sensitive
to the resummation effects beyond next-to-leading order.
 
The remainder of this paper is as follows.  In section 2 we discuss
the BFKL solution and its truncation to ${\cal O}(\alpha_s^3)$.  For
illustrative purposes we also show how this second cross section is
obtained from the
${\cal O}(\alpha_s^3)$ tree-level $2\rightarrow3$ cross section in the
limit of large $y$.  In section 3 we discuss the infrared singularities that
occur at ${\cal O}(\alpha_s^3)$ in order to further elucidate the
approximations used in the BFKL analysis.  We also observe the dependence on
the jet cone-size as a function of $y$.  
In section 4 we compare the BFKL solution
truncated to ${\cal O}(\alpha_s^3)$ with the exact $2\rightarrow3$
${\cal O}(\alpha_s^3)$ cross section.  We isolate the primary contribution to
the discrepancy between the two as arising from approximations used in the
parton distribution functions.  In section 5 we introduce an ``effective
rapidity'' variable to use in the BFKL formula which accounts for this
difference.  In particular, the effective rapidity is defined so that the
truncation of the BFKL solution to ${\cal O}(\alpha_s^3)$ equals the exact
$2\rightarrow3$ ${\cal O}(\alpha_s^3)$ cross section.  We use this modified
BFKL cross section to study the various dijet distributions, which show less
dramatic, but still noticeable, effects than in ref.~\cite{DDS}.
In section 6 we present our conclusions.
 
\section{The minijet resummation and ${\cal O}(\alpha_s^3)$ cross
section at large rapidities}
 
We are interested in the semi-inclusive production of two jets in hard
QCD scattering.  For definiteness we will consider the scattering
process $p_A\bar p_B \rightarrow j_1 j_2 + X$
such as at the Tevatron, but the same analysis can also be
applied to photoproduction at HERA\cite{future}.  We describe the two
partonic tagging jets by their transverse momenta and rapidities
$(\vec{p}_{1\perp}, y_1)$ and $(\vec{p}_{2\perp},y_2)$, where
we always take
$y_1>y_2$.  For large rapidity intervals, $y=y_1-y_2$, the cross
section for this process can be written
\begin{equation}
{d\sigma_0\over dp_{1\perp}^2 dp_{2\perp}^2 d\phi dy_1 dy_2}\
=\ x^0_Ax^0_B\,f_{\rm eff}(x^0_A,\mu^2)f_{\rm eff}(x^0_B,\mu^2)\,
{d\hat\sigma_{gg}\over dp_{1\perp}^2 dp_{2\perp}^2 d\phi}\ ,
\label{general}
\end{equation}
where the parton momentum fractions are dominated by the contribution
from the two tagging jets
\begin{eqnarray}
x^0_A &=& {p_{1\perp} e^{y_1} \over\sqrt{s}}\nonumber\\
x^0_B &=& {p_{2\perp} e^{-y_2} \over\sqrt{s}},\label{pmf}
\end{eqnarray}
and $\mu$ is the factorization/renormalization scale.
In this limit the amplitude is dominated by $gg$, $qg$, and $qq$
scattering diagrams with gluon-exchange in the $t$-channel.  The
relative magnitude of the different subprocesses is fixed by the color
strength of the respective jet-production vertices, so it
suffices to consider only $gg$ scattering and to include the other
subprocesses by means of the effective parton distribution
function\cite{CM}
\begin{equation}
f_{\rm eff}(x,\mu^2) = G(x,\mu^2) + {C_F\over C_A}\sum_f
[Q_f(x,\mu^2) + \bar Q_f(x,\mu^2)]\ , \label{feff}
\end{equation}
In (\ref{feff}) the sum is over the quark flavors, $C_A=N_c=3$ is the
Casimir operator
of the adjoint representation and $C_F = (N_c^2-1)/2N_c = 4/3$ is the one
of the fundamental representation.
 
The higher-order corrections to the $gg$ subprocess cross section in
(\ref{general}) can be expressed via the solution of the BFKL
equation\cite{BFKL}, which is an all-order resummation in
$\alpha_s$ of the leading powers of the rapidity interval
\begin{equation}
 {d\hat\sigma_{gg}\over dp_{1\perp}^2 dp_{2\perp}^2 d\phi} = {C_A^2\alpha_s^2
\over 4\pi p_{1\perp}^3 \, p_{2\perp}^3}
\sum_n e^{in(\phi-\pi)} \int_0^{\infty} d\nu e^{\omega(n,\nu)\, y}
\cos\left(\nu \, \ln{p_{1\perp}^2 \over p_{2\perp}^2} \right),
\label{mini}
\end{equation}
with
\begin{equation}
\omega(n,\nu) = {2 C_A \alpha_s \over\pi} \bigl[ \psi(1) -
 {\rm Re}\,\psi ({|n|+1\over 2} +i\nu) \bigr],
\label{eigen}
\end{equation}
and $\psi$ the logarithmic derivative of the Gamma function.
Eq.~(\ref{mini}) can be expanded order by order in $\alpha_s$ and compared
with a
fixed-order calculation of dijet production at the same order of $\alpha_s$,
in the large rapidity limit\cite{stir}.
By expanding the exponential in
(\ref{mini}) to zeroeth order in $\alpha_s$ we obtain the tree-level
large-$y$ cross section
\begin{equation}
 {d\hat\sigma_{gg}^{(0)}\over dp_{1\perp}^2 dp_{2\perp}^2 d\phi} =
 {\pi C_A^2\alpha_s^2
\over 2 p_{1\perp}^4 }\,\delta(p_{1\perp}^2-p_{2\perp}^2)\,
\delta(\phi-\pi)\ .
\label{tree}
\end{equation}

At ${\cal O}(\alpha_s)$ in the exponential we obtain
\begin{eqnarray}
{d\hat\sigma_{gg}^{(1)}\over dp_{1\perp}^2 dp_{2\perp}^2
d\phi}&=&{C_A^2\alpha_s^2
\over 4\pi p_{1\perp}^3 \, p_{2\perp}^3}
\sum_n e^{in(\phi-\pi)}\label{primo}\\
&&\times\,\int_0^{\infty} d\nu {2 C_A \alpha_s \over\pi} y
\left[ \psi(1) - {\rm Re}\,\psi \left({|n|+1\over 2} +i\nu \right)
\right]
\cos\left(\nu \, \ln{p_{1\perp}^2 \over p_{2\perp}^2} \right)\ .
\nonumber
\end{eqnarray}
When $p_{1\perp} \ne p_{2\perp}$ we can integrate out $\nu$ and sum
explicitly over $n$, by using the integral representation of the $\psi$
function
\begin{equation}
\psi(z) = \int_0^1 dx\ {1-x^{z-1} \over 1-x}\ -\ \gamma
\label{psif}
\end{equation}
with the Euler constant $\gamma=.577215\dots$. We obtain
\begin{equation}
{d\hat\sigma_{gg}^{(1)}\over dp_{1\perp}^2 dp_{2\perp}^2 d\phi} =
{C_A^2\alpha_s^2
\over 4\pi p_{1\perp}^2 \, p_{2\perp}^2}\ {C_A\alpha_s y \over
p_{1\perp}^2 + p_{2\perp}^2 +2p_{1\perp}p_{2\perp}\cos{\phi}}.
\label{explic}
\end{equation}
 
When $p_{1\perp}\sim p_{2\perp}$ the cross section (\ref{primo})
is dominated by configurations where the third parton is soft.
These infrared singularities are regulated by the BFKL solution,
which includes both real and virtual corrections.  To see how this
occurs we can integrate (\ref{primo}) over $p_{2\perp}^2$
in the interval defined
by $|p_{1\perp}^2-p_{2\perp}^2|<\epsilon p_{1\perp}^2$, for $\epsilon$
sufficiently small.  The integrals over $\nu$ and $x$ can be performed,
and the series in $n$ can be summed, giving a finite answer:
\begin{equation}
{d\hat\sigma_{gg}^{(1)}\over dp_{1\perp}^2d\phi} =
{\pi C_A^2\alpha_s^2 \over 2
p_{1\perp}^4}\ \delta(\phi-\pi)\ {2 C_A \alpha_s \over\pi}
y \ln{\epsilon},
\label{virt}
\end{equation}
up to terms of ${\cal O}(\epsilon)$.  For sufficiently small $\epsilon$
this configuration is indistinguishable from a configuration
with only two final state partons.  Thus, for
$|p_{1\perp}^2-p_{2\perp}^2|<\epsilon p_{1\perp}^2$ we can write the
cross section to ${\cal O}(\alpha^3)$ as
\begin{equation}
 {d\hat\sigma_{gg}^{(0+1)}\over dp_{1\perp}^2 dp_{2\perp}^2 d\phi} =
 {\pi C_A^2\alpha_s^2
\over 2 p_{1\perp}^4 }\,\delta(p_{1\perp}^2-p_{2\perp}^2)\,
\delta(\phi-\pi)\ \biggl(1\ +\ {2 C_A \alpha_s \over\pi} y
\ln{\epsilon}\biggr)\ .
\label{nlo}
\end{equation}
For $|p_{1\perp}^2-p_{2\perp}^2|>\epsilon p_{1\perp}^2$ we can use
equation (\ref{explic}).  Combining these two formulae (\ref{nlo}) and
(\ref{explic}), the dependence on the unphysical variable $\epsilon$
will vanish
in any inclusive process integrated over a range of momenta, for
small enough $\epsilon$.
 
It is also informative to
derive the $2\rightarrow3$ cross section (\ref{explic})
by taking the large-rapidity limit of the O($\alpha_s$) real
corrections to gluon-gluon scattering, computed in the
conventional way\cite{exact}. First, note that in the large rapidity limit
the leading contribution is given by the $gg\rightarrow ggg$ subprocess,
the diagrams with final-state quarks being subleading. The squared scattering
amplitude for this, summed (averaged) over final (initial) colors and
helicities, is given in ref.\cite{pt} as
\begin{equation}
|M|^2 \,=\,4\,(\pi\alpha_sC_A)^3 \, \sum_{i>j} \, s_{ij}^4 \,
\sum_{[A,1,2,3,B]'} \, {1 \over s_{A1} s_{12} s_{23} s_{3B} s_{AB}},
\label{parke}
\end{equation}
with i,j~=~A,1,2,3,B, and with the second sum over the noncyclic permutations
of the set [A,1,2,3,B].  The kinematic invariants are defined here by
\begin{eqnarray}
s_{AB} &=& \hat s = x_A x_B s = \sum_{i,j=1}^n p_{i,\perp} p_{j,\perp}
e^{y_i-y_j} \nonumber\\ s_{A,i} &=& -\sum_{j=1}^n p_{i,\perp} p_{j,\perp}
e^{-(y_i-y_j)} \label{inv}\\ s_{B,i} &=& -\sum_{j=1}^n p_{i,\perp} p_{j,\perp}
e^{y_i-y_j} \nonumber\\ s_{ij} &=& 2 p_{i,\perp} p_{j,\perp}
\left[\cosh (y_i-y_j) - \cos (\phi_i-\phi_j) \right]. \nonumber
\end{eqnarray}
In the large rapidity limit eq.~(\ref{parke})
becomes\cite{vd}
\begin{equation}
|M|^2 \,=\,128\,(\pi\alpha_sC_A)^3 \,{{\hat s}^2 \over
p_{1\perp}^2 \, p_{2\perp}^2 \, p_{3\perp}^2}. \label{ampl}
\end{equation}
The large rapidity limit of the phase space for
three-particle production is
\begin{equation}
d\Pi_3 \,=\, {1 \over 2{\hat s}} {dy_{3} \over 4\pi} \left( \prod_{i=1}^3
{d^2p_{i\perp} \over (2\pi)^2} \right) (2\pi)^2 \delta^2\left(\sum_{i=1}^3
p_{i\perp}\right), \label{phase}
\end{equation}
where we have fixed the rapidities of the two gluons at largest and smallest
rapidity and $y_{3}$ is the rapidity of the third gluon.
Using the amplitude~(\ref{ampl}), the phase
space~(\ref{phase}) and the appropriate flux factor, we obtain for the dijet
production cross section $d\hat\sigma_{gg}/ dp_{1\perp}^2 dp_{2\perp}^2 d\phi$
the same expression as the one obtained in eq.~(\ref{explic}) via the
expansion of the BFKL solution.  Note that the overall factor of $y$ in
eq.~(\ref{explic}) comes from the integration of the rapidity of the third
final-state gluon over the interval spanned by the tagging gluons.
 
\section{Collinear singularities in dijet production}
 
Before entering the details of a comparison between the BFKL resummation and
the complete ${\cal O}(\alpha_s^3)$ result, it is useful to consider the
infrared singularities that occur in the two approximations to dijet
production.  These singularities depend on the geometry
of the event and thus are sensitive to how dijet production is defined.
In the next-to-leading order (NLO) approximation
the phase space integration over the kinematic variables of
the third (unresolved) parton generates infrared singularities
when it becomes soft and collinear singularities when it
becomes collinear either with the initial-state partons or with
the other final-state partons.  In contrast, the BFKL approximation only
contains the effects of soft gluons.  Because of the explicit
rapidity-ordering, the unresolved partons are bound to lie
within the rapidity interval defined by the tagging jets, and so
can never become collinear with the initial-state partons\footnote{Of course,
it is always possible for a parton or jet to be
produced outside of the rapidity interval of the tagging jets,
but to go undetected because it has too small $p_{i\perp}$ or too large
$|y_i|$.  In the context of the BFKL approximation this jet would be
considered part of the evolution of the structure functions,
and would not be included in the calculation of the parton subprocess.}.
In addition, in the large rapidity limit the region of phase space where
the third parton becomes collinear with either of the other final-state partons
gives a nonleading contribution.
 
Let us examine this last point further.  First we recall
the standard definition of a jet as consisting of all the partons lying
within a circle of radius $R_{cut}=[(y_J-y)^2+(\phi_J-\phi)^2]^{1/2}$ from
the jet axis $(y_J,\phi_J)$ in the plane defined by rapidity and
azimuthal angle ({\it i.e.}, the Lego plot). Then we look at the dependence
on the jet
cone size $R_{cut}$ of the ${\cal O}(\alpha_s^3)$ corrections to
dijet production. Namely, we consider in Fig. 1 the contribution of the
three-jet configurations, computed through the 2$\rightarrow$3 parton
amplitudes, to the cross section
\begin{equation}
{d\sigma\over dy\,d{\bar y}\,dp_{1\perp}\,dp_{2\perp}}
=\ \int_{y_2}^{y_1}dy_3\int d\phi\sum_{ij} x_A x_B\,f_{i/A}(x_A,\mu^2) 
f_{j/B}(x_B,\mu^2)\,
{d\hat\sigma_{ij}\over dp_{1\perp} dp_{2\perp}dy_3d\phi}\ ,
\label{excross}
\end{equation}
as a function of the jet cone size $R_{cut}$ at different values of the 
rapidity interval $y$. $f_{i(j)} = Q,\bar Q, G$ labels 
the distribution function of the parton species and flavor 
$i$($j$) = $q,\bar q, g$ inside hadron $A$($B$).
We include all parton subprocesses\cite{exact},
and use the exact values of the parton momentum fractions
\begin{eqnarray}
x_A &=& {p_{1\perp} e^{y_1} + p_{2\perp} e^{y_2} + p_{3\perp} e^{y_3}\over
\sqrt{s}}\nonumber\\
x_B &=& {p_{1\perp} e^{-y_1} + p_{2\perp} e^{-y_2} + p_{3\perp} e^{-y_3}
\over\sqrt{s}}.\label{expmf}
\end{eqnarray}
As in our previous analysis\cite{DDS} we set the
rapidity boost $\bar y = (y_1+y_2)/2=0$, since we are mainly interested
in the behavior of the parton subprocess, which does not depend
on $\bar y$.  We fix the transverse momenta of the tagging jets at
different values, $p_{1\perp}$=~20~GeV and $p_{2\perp}$=~50~GeV, in order
to ensure that the third parton cannot become soft.
Furthermore, collinear configurations where the distance $R$ between two of
the partons on the Lego plot is smaller than $R_{cut}$ are discarded, because
the resulting ``combined'' jet would be back-to-back with the remaining parton 
and have the same $p_{\perp}$ as it.
Finally, in the spirit of the BFKL approximation, we restrict the
rapidity of the third parton to lie between the two tagging jets.  Note that
this last approximation differs from the full NLO treatment which includes
initial-state collinear radiation and uses the Altarelli-Parisi subtraction
to handle the divergences.  Therefore, in this calculation and in all BFKL
calculations, the parton density functions should be treated at leading order
(LO) and will have all the associated LO factorization-scale dependence.
In all of our plots we
use the LO CTEQ parton distribution functions\cite{cteq} with
the renormalization and factorization scales set to the geometric mean of the
transverse momenta of the tagging jets $\mu^2 = p_{1\perp} p_{2\perp}$.
As Fig. 1 shows, the dependence of the dijet production on the jet cone size
$R_{cut}$ decreases as $y$ is increased. At $y=6$ the tagging jets are so
widely separated in rapidity that there is almost no dependence on $R_{cut}$
at all. This confirms that the phase space region where two final-state
partons are collinear is subleading at large rapidities,
as suggested by the BFKL theory.
 
\section{The BFKL and the ${\cal O}(\alpha_s^3)$ corrections to the
$p_{\perp}$ distributions}
 
In section 2 we have shown that the truncated ${\cal O}(\alpha_s^3)$ BFKL
contribution to dijet production is the large rapidity limit of
the complete ${\cal O}(\alpha_s^3)$ corrections. Here we want to see how
well this works in practice in the jet-jet $p_{\perp}$ correlation case
examined in ref.\cite{DDS}. In Fig.~2 we consider
the contribution of the three-jet amplitudes to the $p_{\perp}$ correlation
$d\sigma/dy\,d{\bar y}\,dp_{1\perp}\,dp_{2\perp}$,
plotted as a function of the transverse momentum
$p_{1\perp}$, at a fixed value of $p_{2\perp}$=~50~GeV and at
$y$~=~2 and 6. The customary value for the jet cone size $R_{cut}$=~0.7 has
been used. As in fig.~1, configurations where the distance $R$ between two
partons on the
Lego plot is smaller than $R_{cut}$ are discarded, since they would be
counted as
a two-jet event with $p_{1\perp}=p_{2\perp}$.  We go from the exact
configurations to the large-$y$ approximation to them in three steps.
The dashed curves are computed through the exact
2$\rightarrow$3 parton amplitudes\cite{exact} and kinematics (\ref{expmf});
the dotted curves through the large-$y$ parton amplitudes
(\ref{ampl}) and the exact kinematics (\ref{expmf});
the solid curves through the large-$y$ parton amplitudes (\ref{ampl}) and
kinematics (\ref{pmf}). As the plots show, the error in using the large-$y$
approximation grows with the imbalance in transverse momentum of the tagging
jets. While at small $y$'s the error is distributed between the
approximation on the amplitudes and the one on the parton distribution
functions, at large $y$'s most of the error comes in using the
large-$y$ approximation in the parton distribution functions.
We have also made this comparison for the larger rapidity intervals obtainable
at the CERN Large Hadron Collider (LHC) and have found that this discrepancy,
although smaller, is not insignificant.
 
This discrepancy can be understood by recalling that the rapidity $y_3$ of
the third jet is integrated over the full range of the interval from $y_2$ to
$y_1$.  If we neglect its contribution to the momentum fractions as in
(\ref{pmf}), this just multiplies the cross section by a factor of $y$ as in
eq.~(\ref{explic}).  However, eq.~(\ref{pmf}) can be a bad approximation to
the exact kinematics (\ref{expmf}) over much of the integral if $p_{3\perp}$
is not small.  For $y_3$ near the extremes, using the exact kinematics in
the parton distribution functions produces a large suppression, so that the
``effective'' rapidity range of $y_3$ is reduced substantially.  The
truncated ${\cal O}(\alpha_s^3)$ BFKL (solid) curve neglects this effect, and
so it overestimates the cross section.  Note, however, that
near $p_{1\perp}=p_{2\perp}$ the transverse momentum of the third parton is
small, so its contribution to the $x$'s in (\ref{expmf}) can be safely
neglected.
 
\section{An effective rapidity interval}
 
We have just seen in the previous section that the large-$y$ approximation
used in the BFKL
resummation seriously overestimates the cross section when the two tagging
jets have unequal energies, even for rapidity intervals as large as $y=6$.
This occurs because the large-$y$ cross section (\ref{explic}) assumes that
the third (minijet) parton can be produced anywhere within the rapidity
interval $[y_2,y_1]$ with equal probability, whereas in the full
$2\rightarrow3$ cross section the probability is highly suppressed by the
structure functions when the third jet strays too far from the center of this
interval.  In this section we will attempt to fix this problem by including
this suppression effect directly into the BFKL equation.
 
The BFKL solution as presented in equation (\ref{mini}) is an all orders
resummation in $\alpha_s y$ where $y$ is the kinematic rapidity interval
defined by the tagging jets.  However, to be completely precise, the rapidity
variable which is resummed by BFKL
is only defined up to transformations $y\rightarrow y+X$ where $X$
is subleading at large rapidities.  This is analogous to a change in the
scale $Q^2$ in the standard $\alpha_s \ln(Q^2)$ resummation, and in the same
way that one chooses a physical value of $Q^2$ to lessen the effects of
nonleading terms in this resummation, one can choose a more physical rapidity
variable in the BFKL resummation.  From the results of the previous section
we see that a better rapidity variable would be one that reflects the range
in rapidity spanned by the minijets, which is typically less than the
kinematic rapidity interval $y$.  Let us define an ``effective rapidity''
$\hat y(n,p_{1\perp},p_{2\perp},\bar y, y)$ by
\begin{equation}
\hat y\ \equiv\ y\ {
\displaystyle \int d\phi\, \cos(n\phi)\ \big(d\sigma/dyd\bar y
dp_{1\perp}dp_{2\perp}d\phi\big) 
\over
\displaystyle\int d\phi\, \cos(n\phi)\ \big(
d\sigma_0/dyd\bar ydp_{1\perp}dp_{2\perp}d\phi\big)}\ ,\label{effecty}
\end{equation}
where $n$ is the fourier series index of eq.~(\ref{mini}).
The cross section in the numerator is that of eq.~(\ref{excross}) and is 
computed 
using the exact kinematics (\ref{expmf}), while the cross section in the 
denominator is that of eq.~(\ref{general}) and is computed using the large-$y$ 
kinematics (\ref{pmf}).  The denominator can easily
be computed analytically using the large-$y$ solution
(\ref{explic}).  Note that $\hat
y$ is defined so that if we replace $y\rightarrow\hat y$ in the BFKL solution
(\ref{mini}) and truncate to ${\cal O}(\alpha_s^3)$ we recover the exact
$2\rightarrow3$ cross section.  Also note that asymptotically for large $y$,
the difference $y\!-\!\hat y$ is nonleading.  Thus, we can use $\hat y$ in
(\ref{mini}) and obtain a quantitatively more reliable solution.
 
In Fig.~3 we plot $\hat y$ as a function of $p_{1\perp}$ for $n=0$,
$p_{2\perp}=50$ GeV, $\bar y=0$, and $y=2,3,4,5,6$ with $R_{cut}=0.7$.
For $y=2$ and 6 this is just given by the ratio of the dashed curves to the
solid curves in Fig.~2, multiplied by $y$.  Near $p_{1\perp}=p_{2\perp}$,
$\hat y$ approaches
the kinematic rapidity $y$, especially for large $y$, but it falls
quickly as the two jets move apart in transverse momentum.  Thus, we would
expect the effects of the resummation to be most important near
$p_{1\perp}=p_{2\perp}$.  We can see this clearly in the jet-jet
$p_{\perp}$ correlation plot of Fig.~4, where
we compare the exact $2\rightarrow3$ ${\cal O}(\alpha_s^3)$ cross section
(dashes) with the BFKL resummation using $\hat y$ (solid).  As in Fig.~2 we
fix $p_{2\perp}=50$ GeV, $\bar y=0$, and plot as a function of $p_{1\perp}$
for $y=2$ and 6, using a jet cone size of $R_{cut}=0.7$.  In this and in the
plots that follow, $\hat y$ is fixed
at $n=0$ by the integration over the azimuthal angle $\phi$. As expected,
for $y=2$ there is little difference between the two approximations. For $y=6$
the higher orders of the BFKL resummation are very important near
$p_{1\perp}=p_{2\perp}$, but have less of an effect when the jets are
mismatched in transverse momentum.
 
Next, in Fig.~5 we plot the transverse momentum distribution
$d\sigma/dy\,d{\bar y}\,dp_{1\perp}$ of jet~1 at $y=4$ and 6 and $\bar y$~=~0.
The dashed curves are computed through the exact Born-level matrix elements
and the solid curves using $\hat y$ in the BFKL resummation and
two different cutoffs for jet~2
transverse momentum, $p_{2\perp min}$=~10~GeV and 20~GeV.
The effect of the radiative corrections is to release the $p_{\perp}$
distribution from the Born-level requirement that
$p_{1\perp}=\, p_{2\perp}$. At a fixed order of $\alpha_s$
and for a given $p_{1\perp}$, jet~2 and the minijets
try to have $p_{\perp}$ as small as possible in order to minimize
their contribution to the $x$'s (\ref{expmf}) and thus maximize
the value of the parton distribution functions. They are only constrained by
the overall transverse momentum conservation.
Eventually when the number of minijets is very
high, or virtually infinite as in the BFKL resummation, the smallest
value of $p_{2\perp}$ is not fixed any more by transverse
momentum conservation but by the minimum $p_{\perp}$ experimentally attainable
\cite{DDS}. This explains the strong sensitivity of the curves in Fig.~5 to
the value of $p_{2\perp min}$\footnote{On the other hand, light-cone momentum
conservation requires that $p_{1\perp} < \sqrt{s} \exp(-y_1)$, so that
at large rapidity intervals
the maximum value of $p_{1\perp}$ does not appreciably
change going from the Born level to higher order corrections.}.
The crossing of the curves with $p_{2\perp min}=20$ GeV below the Born curves 
for small values of $p_{1\perp}$ is a further manifestation of this 
decorrelation in transverse momentum.  For $p_{1\perp}\sim20$ GeV, jet 2 very 
often radiates away enough energy so that $p_{2\perp}<p_{2\perp min}$.  
Thus, events that would have been kept at the Born level are now discarded 
when higher orders are included.
Finally, for sake of comparison, we replot in the dotted curves the 
$p_{\perp}$ distribution computed at $y=4$ using
$y$ in the BFKL resummation, with $p_{2\perp min}$=~10~GeV and 20~GeV
\cite{DDS}.
We note that, using the effective rapidity $\hat y$ rather than the kinematic 
rapidity $y$ in the BFKL resummation, the enhancement in the $p_{\perp}$ 
distribution is considerably reduced.
 
To examine further the kinematic effects on dijet production we plot in
Fig.~6 the $p_{\perp}$ distribution at $|\bar y|=2$ and $y=4$ as a function
of the transverse momentum of the jets at largest ($|y_f|=4$) and
smallest ($|y_c|=0$) 
absolute rapidities, which we call the {\it forward} jet and the {\it
central} jet, respectively.  As in Fig.~5 we use
two different cutoffs $p_{\perp min}=10$~GeV and 20~GeV for the transverse
momentum integrated out. Since we
only have changed $\bar y$ with respect to Fig.~5 the contribution of the
parton subprocess to the $p_{\perp}$ distribution will be the same as in
Fig.~5, but the contribution of the parton distribution functions will change.
The dashed curve is computed through the exact Born-level matrix elements
and the dotted and solid curves using $\hat y$ in the BFKL resummation.
At the Born level
the transverse momentum distributions of the forward and central jets are the
same since $p_{f\perp}=\, p_{c\perp}$, but they become starkly
different when higher orders are included.  The dotted curves are the
distributions of the forward jet. They do not appreciably differ from
the Born-level 
curve because the upper bound $p_{f\perp} < \sqrt{s} \exp(-|y_f|)$
from light-cone momentum conservation
is very restrictive and the phase space of the forward jet
does not basically change going from LO to higher order corrections.
On the other hand the solid curves, which represent the distributions
of the central jet,
show a huge enhancement and very different slopes as compared to
the Born level. To understand this, we have to look at the
kinematics of the higher order corrections. At ${\cal O}(\alpha_s^3)$ it is
possible to let $p_{c\perp}$ grow by taking the third jet at small
absolute rapidity and almost back-to-back with the central jet,
$p_{3\perp}\simeq -p_{c\perp}$ and $y_3\simeq 0$, while taking the
forward jet 
as slim as possible, $p_{f\perp}\simeq p_{\perp min}$.
Then from light-cone momentum conservation
$p_{c\perp max} \simeq \left(\sqrt{s}-p_{\perp min}\exp(|y_f|)\right)/2$,
which is much bigger than the Born-level upper bound.
This picture generalizes to higher orders with the transverse
momentum of the minijets balancing that of the central jet, while the
forward jet is produced near $p_{\perp min}$. 
In order to test this picture
we consider in the dot-dashed curve the transverse
momentum distribution for the Born-level production of two jets in the
central region $|\bar y|=0.5$ and $y$=1. The slope of this curve is similar to
that of the solid curves, suggesting that the $p_{c\perp}$ distribution
is dominated by events with two central jets and one soft
forward jet, where we tag on the forward jet and one of the central
jets.  In the BFKL resummation the second hard central jet is
presumably replaced by a succession of minijets.  
Thus, it is misleading to compare the
BFKL resummation to the Born level (dashed) for the $p_{c\perp}$ 
distribution, since it is dominated by events with more than two jets.
A better comparison in this case would be between the
BFKL resummation and a NLO calculation.
 
\section{Conclusions}

In this paper we have attempted to better grasp the range of validity of the 
BFKL resummation by comparing it with an exact $2\rightarrow3$ ${\cal O}
(\alpha_s^3)$ calculation.  We saw that the dependence on the jet cone-size 
quickly becomes insignificant for moderate rapidities, just as is assumed
in the BFKL formalism.  However, the approximation to the kinematics in
the parton momentum fractions causes a serious error in the BFKL predictions
when the tagging jets are not back-to-back in $p_{\perp}$ and $\phi$.  In
order to account for this error we introduced an effective rapidity $\hat y$
which restricts the phase space of the minijets in such a way that the 
truncation of the BFKL resummation to ${\cal O}(\alpha_s^3)$ agrees with the 
exact $2\rightarrow3$ ${\cal O}(\alpha_s^3)$ calculation.

Using the BFKL resummation with the effective rapidity $\hat y$ we have seen
that the effects on the $p_{\perp}$ spectrum are not as dramatic as we had
predicted previously.  The difficulty in detecting these deviations from
the Born-level computation are compounded by renormalization/factorization
scale ambiguities, which are at least as problematic here as at the Born
level.  Because of the two scales defined by the tagging jets we could let
$\mu^2 = Cp_{1\perp}p_{2\perp}$, $C{\rm max}(p_{1\perp}^2,p_{2\perp}^2)$,
or some other choice where $C$ is some constant of order one.  Because of 
the relatively small deviations of the BFKL resummation with the effective
rapidity $\hat y$ from the Born-level calculation, and the sizeable 
renormalization/factorization scale ambiguities in the BFKL approximation,
we expect that a complete NLO calculation could give a more reliable
estimate to the raw $p_{\perp}$ spectra.  However, we do note that much of the
uncertainties due to the renormalization/factorization scale drop out in
the ratios of cross sections so that, for instance, the predictions of the
ratio of the $p_{1\perp}$ spectra with different $p_{2\perp min}$ cutoffs
are probably reliable to $\sim15\%$.  In addition, the large corrections in 
the $p_\perp$ correlation and $\phi$ correlation plots suggest that an
NLO calculation would be inadequate in these cases, and that the BFKL
resummation using the effective rapidity $\hat y$ should do a better job
here.      
 
\section*{Acknowledgements}
 
We wish to thank Jerry Blazey, Terry Heuring, Soon Yung Jun, Chang Lyong Kim,
David Kosower, Al Mueller, and Paolo Nason for useful discussions.

\newpage

\begin{figure}
\vskip-6cm
\epsfysize=16cm
\centerline{\epsffile{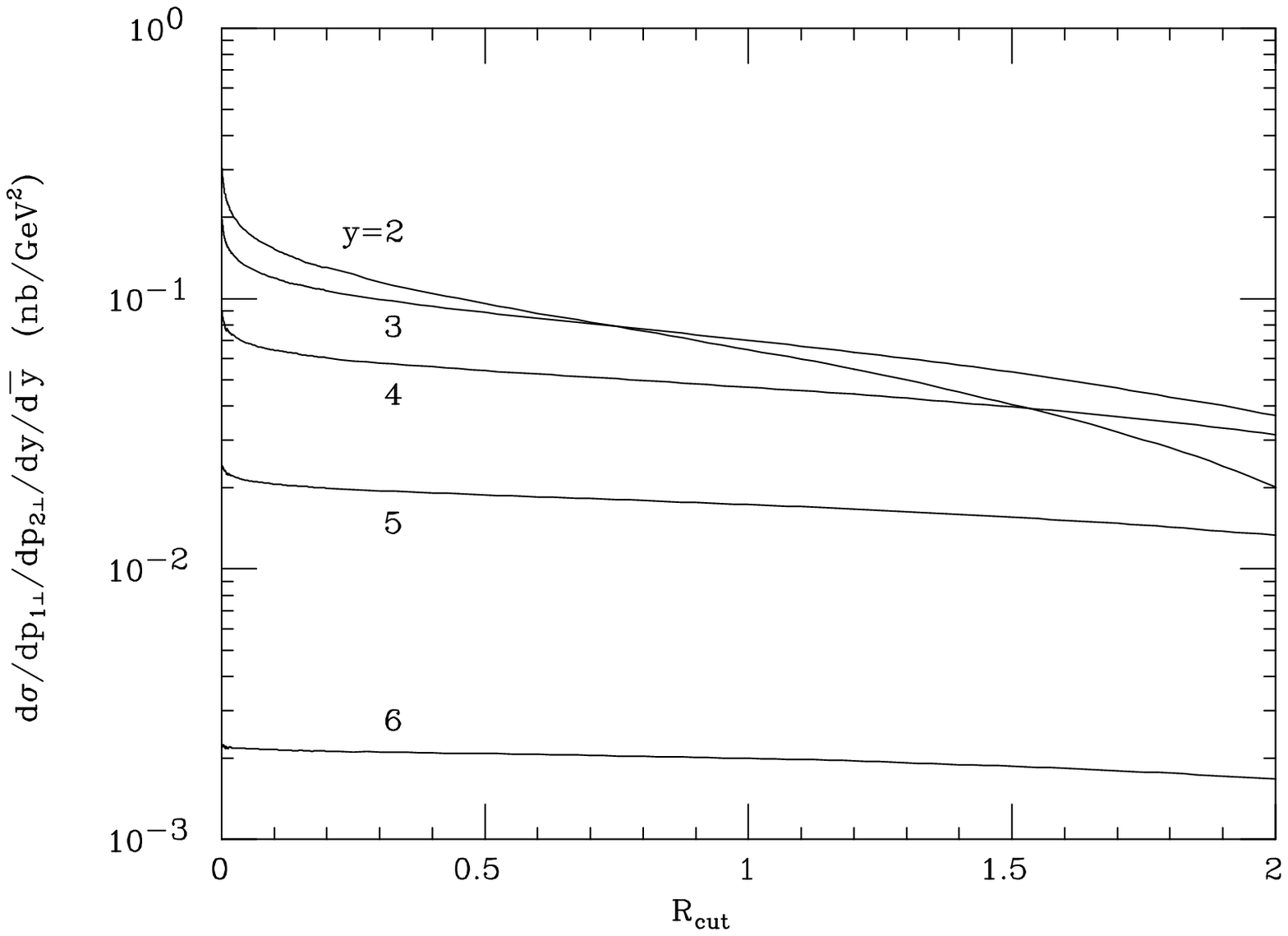}}
\vskip-4cm
\vskip6pt
\baselineskip=12pt
Fig.~1: Dijet production as a function of the jet cone size $R_{cut}$,
at $p_{1,\perp}$~=~20 GeV and $p_{2,\perp}$~=~50 GeV, $\bar y$~=~0 and
$y$~=~2, 3, 4, 5 and 6.
\vskip-3cm
\epsfysize=16cm
\centerline{\epsffile{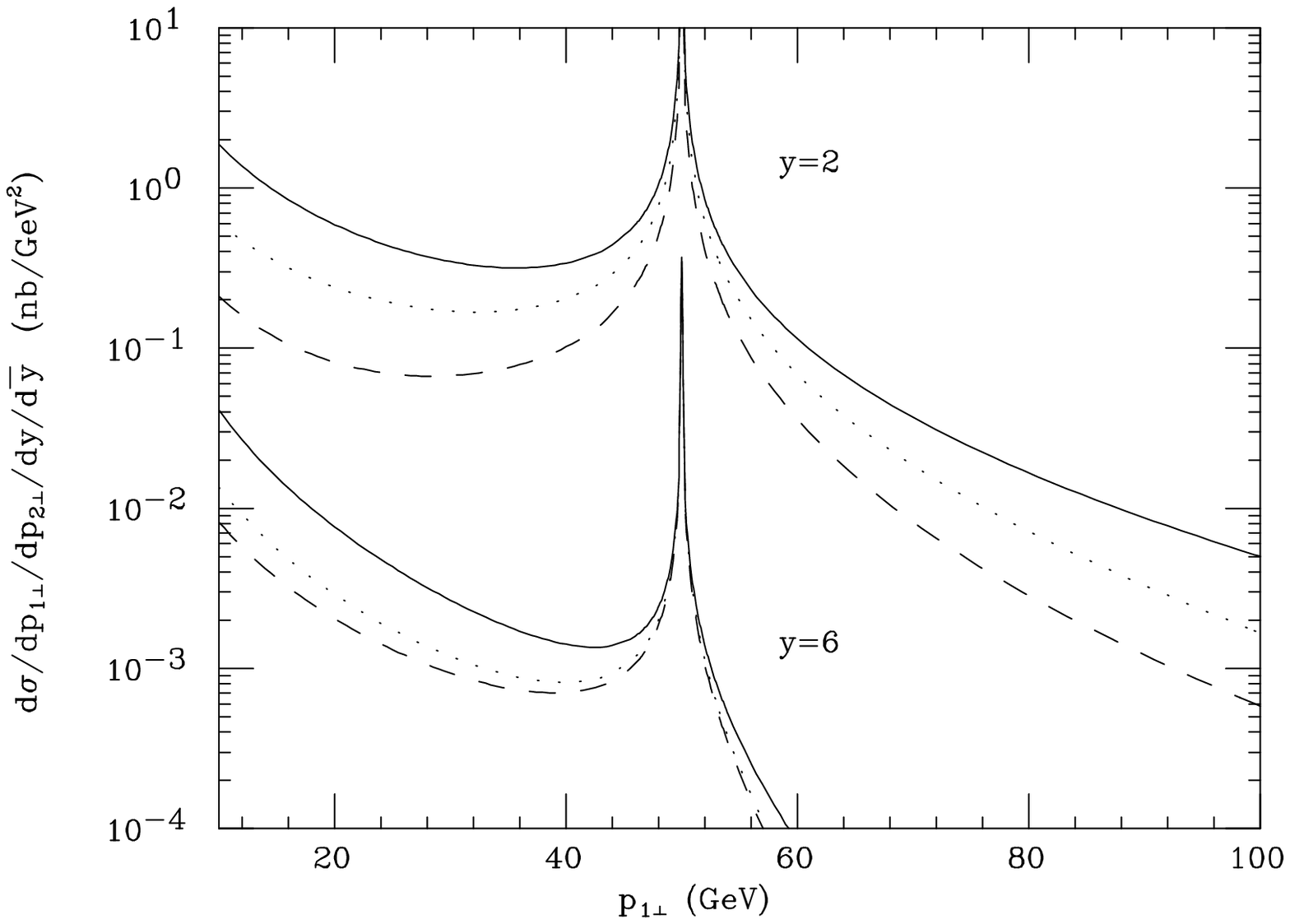}}
\vskip-4cm
\vskip6pt
\baselineskip=12pt
Fig.~2: $p_{\perp}$ distribution of jet 1 with the jet 2 transverse momentum
fixed at 50~GeV, at $\bar y$~=~0 and at $y$~=~2 and 6.
The jet cone size is fixed at 0.7. The dashed curves are computed through the
exact
2$\rightarrow$3 parton amplitudes and kinematics; the dotted curves through
the large-$y$ parton amplitudes and the exact kinematics; the solid curves
through the large-$y$ parton amplitudes and kinematics.
\end{figure}

\newpage

\begin{figure}
\vskip-6cm
\epsfysize=16cm
\centerline{\epsffile{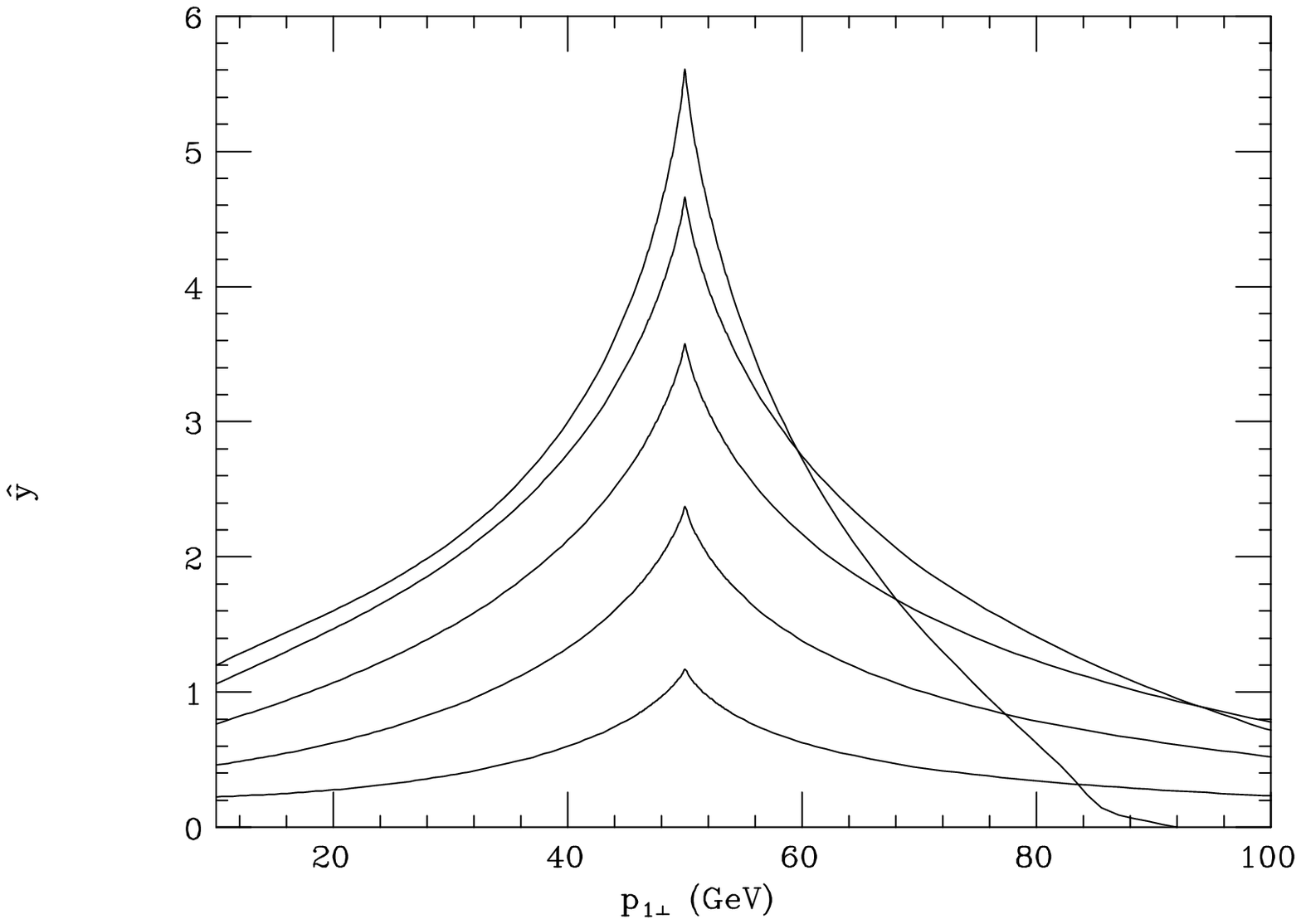}}
\vskip-4cm
\vskip6pt
\baselineskip=12pt
Fig.~3: $\hat y$ as a function of $p_{1\perp}$ for fixed $p_{2\perp}=50$ GeV
and $n=0$, $\bar y$~=~0.  The curves from bottom to top are for $y$~=~2, 3,
4, 5, and 6.  The jet cone size is fixed at 0.7. The upper end of the curve
$y$~=~6 is cut off because $p_{1\perp}$ reaches the kinematic limit.
\vskip-3cm
\epsfysize=16cm
\centerline{\epsffile{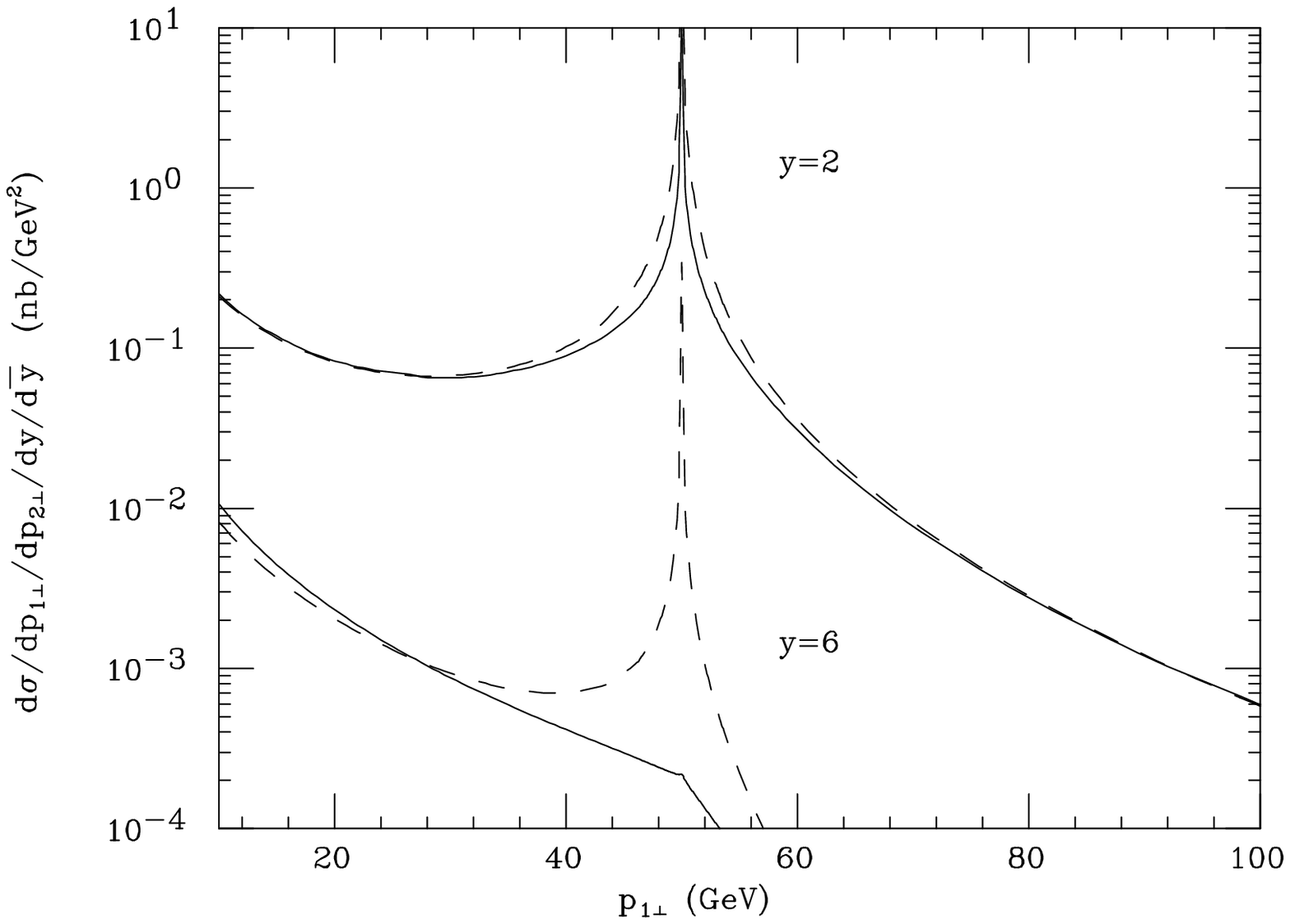}}
\vskip-4cm
\vskip6pt
\baselineskip=12pt
Fig~4: $p_{\perp}$ distribution of jet 1 with the jet 2 transverse momentum
fixed at 50~GeV, at $\bar y$~=~0 and at $y$~=~2 and 6.
The jet cone size is fixed at 0.7. The dashed curves are computed through the
exact 2$\rightarrow$3 parton amplitudes and kinematics; the solid curves are
computed from the full BFKL solution using $\hat y$.
\end{figure}

\newpage

\begin{figure}
\vskip-6.5cm
\epsfysize=16cm
\centerline{\epsffile{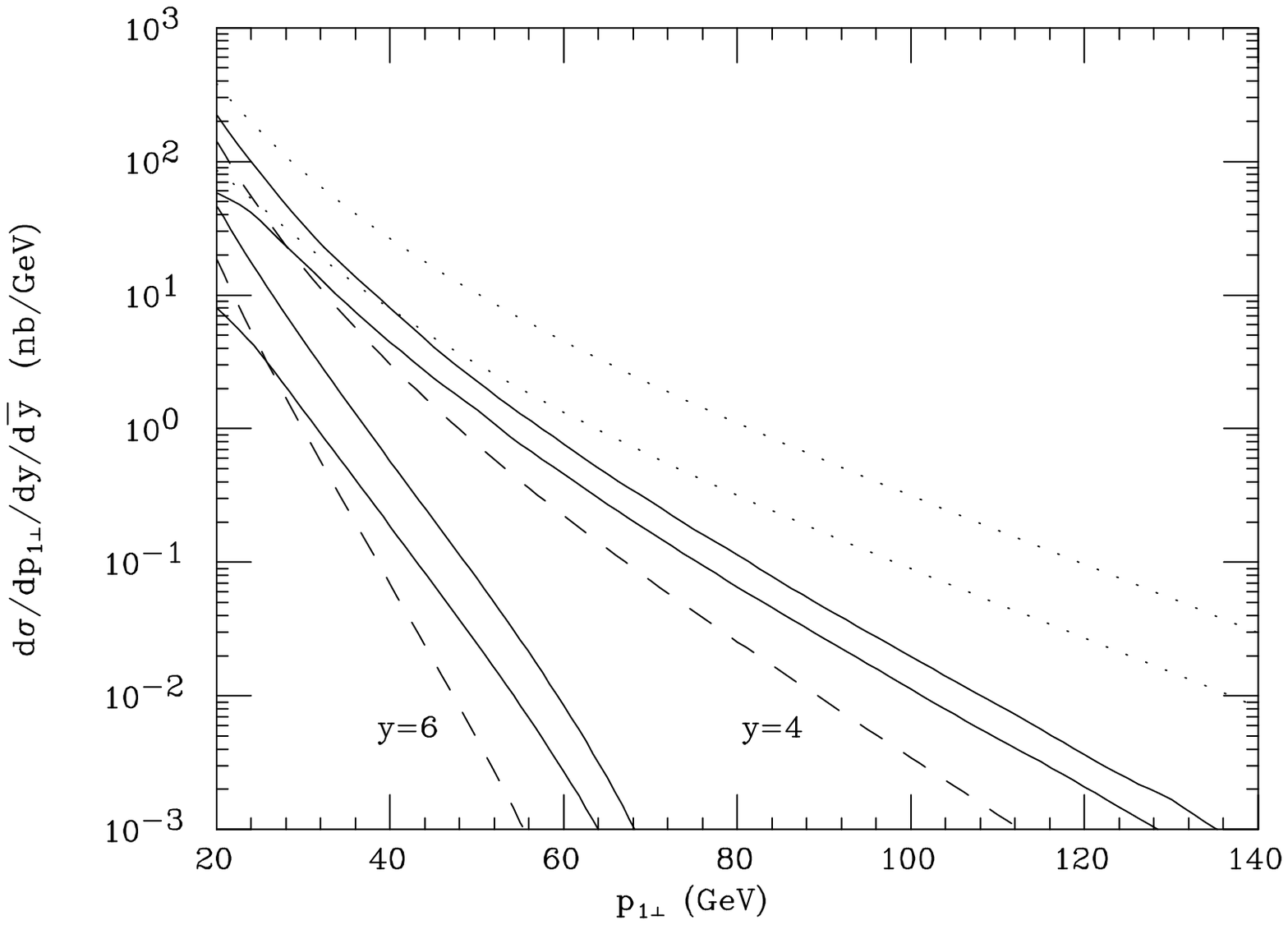}}
\vskip-4.5cm
\vskip6pt
\baselineskip=12pt
Fig.~5: $p_{\perp}$ distribution of jet 1 at $\bar y$~=~0 and $y$~=~4 and 6. 
The dashed curves are the exact Born-level $p_{\perp}$ distributions.
The solid curves are the $p_{\perp}$ distributions computed using $\hat y$
in the BFKL resummation, with two different cutoffs for jet~2,
the upper curve with $p_{2\perp min}$=~10~GeV and 
the lower curve with $p_{2\perp min}$=~20~GeV. The dotted curves 
are the $p_{\perp}$ 
distributions at $y=4$ only, computed using the kinematic rapidity $y$ in the 
BFKL resummation, with $p_{2\perp min}$=~10~GeV and 20~GeV.
\vskip-3cm
\epsfysize=16cm
\centerline{\epsffile{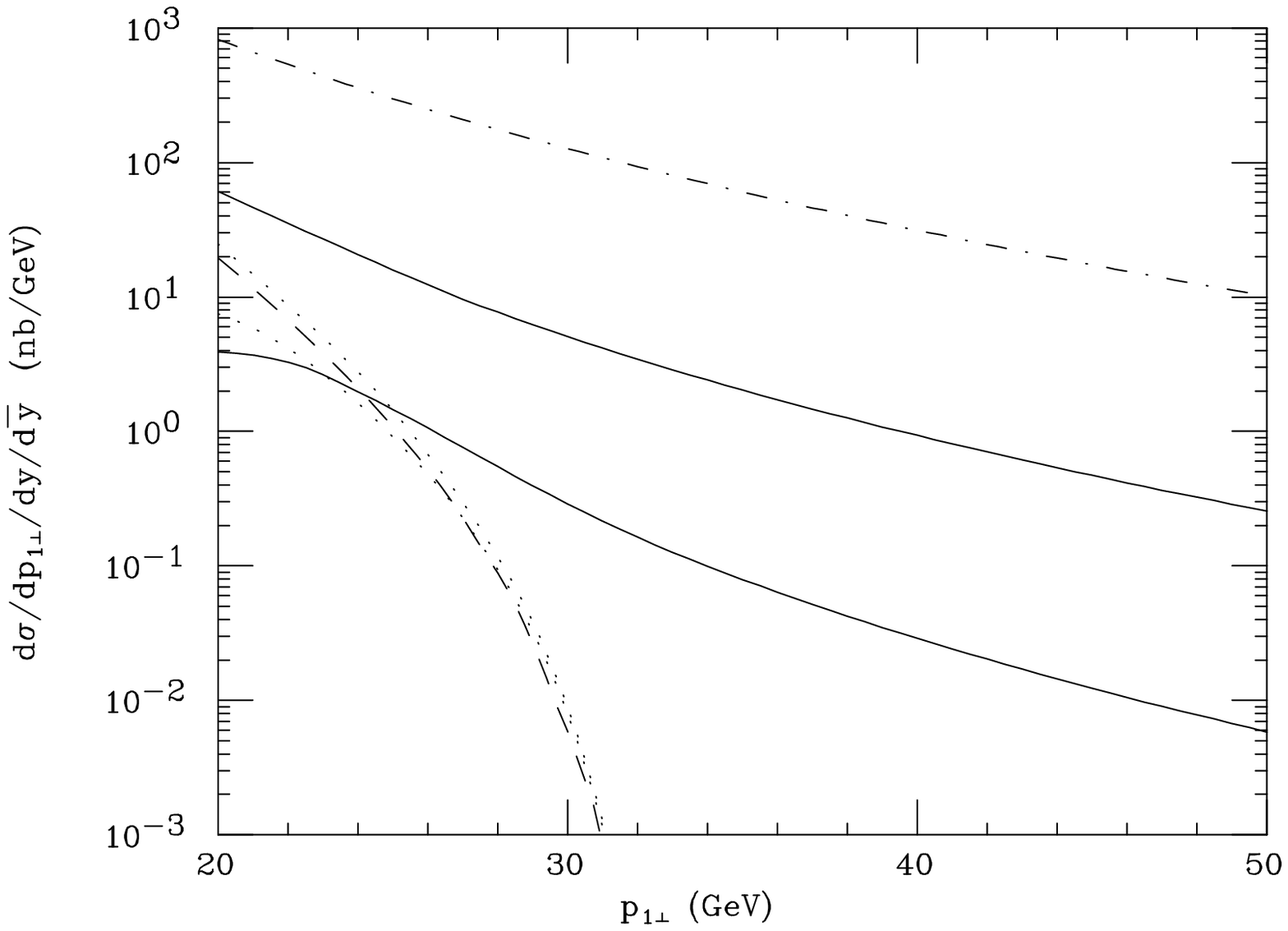}}
\vskip-4.5cm
\vskip6pt
\baselineskip=12pt
Fig.~6: $p_{\perp}$ distribution at $|\bar y|=2$ and $y=4$. The
dashed curve is the exact Born-level $p_{\perp}$ distribution.
The dotted (solid) curves are the $p_{\perp}$ distributions of the forward 
(central) jet computed using $\hat y$ in the BFKL resummation, with two 
different cutoffs for the transverse momentum
integrated out, the upper curve with $p_{\perp min}=10$~GeV and 
the lower curve with $p_{\perp min}=20$~GeV . The dot-dashed curve is
the exact Born-level $p_{\perp}$ distribution at $|\bar y|=0.5$ and $y=1$.
\end{figure}

\end{document}